\documentclass[showpacs,amsfonts,eqsecnum,twocolumn,aps,prd]{revtex4} 

\usepackage{amsmath2000}
\usepackage{epsfig}
\usepackage{bbm} 

\newcommand\NPA{{Nucl. Phys.} A}
\newcommand\NPB{{Nucl. Phys.} B}

\newcommand\PLB{{Phys. Lett.} B}

\newcommand\PRL{Phys. Rev. Lett.}
\newcommand\PRC{{Phys. Rev.} C}
\newcommand\PRD{{Phys. Rev.} D}

\newcommand\JPG{{J. Phys.} G}

\newcommand\IJMPA{{Int. J. Mod. Phys.} A}

\newcommand\CQG{Class. Quant. Grav.}

\newfam\BMath
\font\BMathL=cmmib10 
\font\BMathl=cmmib7
\font\BMathm=cmmib5
\textfont\BMath=\BMathL \scriptfont\BMath=\BMathl
\scriptscriptfont\BMath=\BMathm

\renewcommand\a{\alpha}

\renewcommand\d{\delta}
\newcommand\e{\epsilon}

\newcommand\h{\eta}

\renewcommand\l{\lambda}
\newcommand\m{\mu}
\newcommand\n{\nu}

\newcommand\p{\pi}
\newcommand\s{\sigma}
\renewcommand\t{\tau}

\newcommand\f{\phi}

\renewcommand\exp{\mbox{\rm exp}}  
\newcommand\tr{\mbox{\rm tr}} 

\newcommand\ra{\rightarrow}

\newcommand\Lra{\longrightarrow}

\newcommand{\lan}{\langle}     
\newcommand{\ran}{\rangle}     

\newcommand\del{\partial}

\newcommand{\fxd}{\!\!}          

\newcommand{\nonum}{\nonumber}

\newcommand{\half}{\frac{1}{2}}
\newcommand{\2}{\frac{1}{2}}

\newcommand{\4}{\frac{1}{4}}

\newcommand\be{\begin{equation}}
\newcommand\ee{\end{equation}}
\newcommand\bea{\begin{eqnarray}}
\newcommand\eea{\end{eqnarray}}
\newcommand\bal{\begin{align}}
\newcommand\eal{\end{align}}
\newcommand\bwt{\begin{widetext}}
\newcommand\ewt{\end{widetext}}

\newcommand\ba{\begin{array}}
\newcommand\ea{\end{array}}
\newcommand\bc{\begin{center}}
\newcommand\ec{\end{center}}

\newcommand\eref[1]{Eq.~(\ref{#1})}
\newcommand\etwref[2]{Eqs.~(\ref{#1}) and (\ref{#2})}
\newcommand\ethref[3]{Eqs.~(\ref{#1}), (\ref{#2}) and (\ref{#3})}

\newcommand\bfi{\begin{figure}}
\newcommand\efi{\end{figure}}
\newcommand\bpi[1]{\begin{picture}#1}
\newcommand\epi{\end{picture}}

\def\jou#1#2#3#4{{#1} {\bf #2}, #3 (#4)}

\newcommand\sst{\scriptstyle}

\newcommand\thtw{\frac{3}{2}}

\newcommand\rttw{\sqrt{2}}
\newcommand\rtsx{\sqrt{6}}

\newcommand\ve{\varepsilon} 

\newcommand\BR{{\mathbb R}}
\newcommand\BZ{{\mathbb Z}}

\newcommand\bfa{\text{\boldmath $a$}} 
\newcommand\bft{\text{\boldmath $\t$}} 

\newcommand\ha{{\hat a}}
\newcommand\hp{{\hat \p}}
\newcommand\hI{{\hat I}} 
\newcommand\hJ{{\hat J}} 
\newcommand\hL{{\hat L}} 
\newcommand\hP{{\hat \Pi}} 
\newcommand\hR{{\hat R}}

\newcommand\shalf{\mbox{$\frac{1}{2}$}}

\def\bem{\begin{pmatrix}}
\def\eem{\end{pmatrix}} 

\newcommand\bse{\begin{subequations}}
\newcommand\ese{\end{subequations}} 

\newcommand\phm{\phantom{-}}

\begin{document}

\title{
Equivalence of Classical Skyrmions and Coherent States of Baryons
\\ 
I. Constrained Quantization on the $SU(2)$ and $SO(3)$ manifolds  
} 

\author{S.M.H. Wong} 
\affiliation{Department of Physics, The Ohio State University, Columbus, 
Ohio 43210} 

\begin{abstract} 

In the Skyrme model, the Lagrangian can be quantized in several ways
using the collective coordinate approach. Not all of which produce
quantum states that can be interpreted as physical particles. 
For example the $SU(2)$ collective coordinate approach produces both 
integral and half-integral spin and isospin states. Only half of these 
are the physical baryons. Less well known is the fact that it is equally 
possible to quantize the system using the $SO(3)$ collective coordinates. 
This produces only unphysical integral spin and isospin states. 
To fulfill the goal of being able to express a classical skyrmion as 
a coherent state of baryons directly in terms of the baryon states,   
surprisingly a combination of both collective coordinate approaches  
is required. To prepare for the subsequent application to skyrmion 
formation through disoriented chiral condensates in heavy ion collisions, 
the Skyrme model is rigorously quantized using the Dirac prescription
for constrained systems. This is shown for both the $SU(2)$ as well
as the $SO(3)$ collective coordinate approach. 

\end{abstract}  

\date{\today} 
 
\null 


\maketitle 

\section{Introduction} 
\label{s:intro}

Four decades ago contrary to the popular trend at the time to build 
theories in terms of fermions, Skyrme succeeded in building a model with 
a Lagrangian density based on meson fields. Not that it was particularly 
difficult to construct meson field theories but rather this 
one now known as the Skyrme model has the remarkable feature that 
fermions can come out of boson fields in the form of topological solitons 
\cite{sk1,sk2}. Those of the Skyrme model are known as skyrmions. The 
topological charge or winding number $B$ of skyrmions was correctly identified 
by Skyrme as the baryon number and the $B=1$ skyrmion was conjectured 
to be a fermion \cite{sk1,sk2}. It was only until much later through the method 
of Ref. \cite{gw} that the topological charge was worked out to be equal to the 
baryon number \cite{bnrs1,bnrs2}. Ref. \cite{w2} was also able to rewrite the
latter in the form of the winding number and thus made the connection.
As to the conjecture of a $B=1$ skyrmion is a fermion, this can only be
settled by considering three or more flavors of quarks when the Wess-Zumino
effective action \cite{wz} was introduced to break an unphysical symmetry
in the Skyrme model \cite{w1}. Only with this action can one show that
under a rotation by an angle of $2\p$, a skyrmion acquires a phase change
of $\exp (iN \p)$ \cite{w2}. A $B=1$ skyrmion is therefore a fermion and a
baryon. These distinctive works eliminated any doubt on the species of particle
that a skyrmion belongs. Therefore the Skyrme model in the $B=1$ sector
became a non-trivial model of baryon or nucleon. Many further works have been 
produced using this model and many applications were found. For example 
for studying the properties such as masses, magnetic moments, sizes,
interactions, structure, and form factors of nucleons and baryons
\cite{anw,jr,rgb,gu,wk,kvw}. Other remarkable points about the Skyrme model
are that it can explain the fact that most spin in a proton is not carried by
the valence quarks and that there is the existence of the $q\bar q$ sea inside
the nucleon whereas the more basic non-relativistic quark model completely
fails in this regard \cite{bek,dn}. There are also attempts to go beyond the
two-flavor Skyrme model. This has proven to be not so simple. There are
at least two ways to do this. One is to continue with the collective
coordinate approach initiated in Ref. \cite{anw} by embedding the flavor
$SU(2)$ skyrmion in the higher flavor $SU(N)$ theory \cite{gu,mnp,wall},
the other is to bind a kaon to a flavor $SU(2)$ skyrmion and treating
strange quark as distinct from the light quarks \cite{wk,ck,nnrv,snnr,vk}.
Unfortunately both approaches have their own problems in getting a
reasonable estimates of the baryon mass spectrum. Some reviews of both 
approaches as well as on other applications of the three-flavor 
skyrmions can be found in Ref. \cite{hs,hw}. On two- or more flavor skyrmions 
these have been considered as deuteron or other massive nuclei, one 
can consult for example Ref. \cite{ec}. 

Our interest in the skyrmion stems from an apparently rather remote 
subfield of physics. While most of the works mentioned above were interested 
in the general properties, static or otherwise, of the baryons, we 
are more concerned with the dynamical production of skyrmions in the seemingly
rather unlikely context of heavy ion collisions. The motivation behind this 
is the possibility of observing skyrmion production. This calls for an 
examination of the skyrmion in a very different light. Rather than asking 
the questions of what are the physical properties of the quantum 
states of the Skyrme model, the question we ask ourselves is what exactly 
is the quantum analog of the classical skyrmion solution. If one does not 
know what the classical solution corresponds to in terms of physical 
baryon states, there is little chance for their observation. 

This possibility of skyrmion production in heavy ion collisions was first
raised in Ref. \cite{dg}. Strangely this was at least several years before the
same possibility was considered in jets \cite{ek1,ek2} and reapplied to heavy
ion collisions \cite{ehk,ks}. Within the context of heavy ion collisions
skyrmions are considered as topological defects which might form when the
system passes from the chiral symmetric phase back into the broken phase
just as strings and monopoles might form during the electroweak phase
transition in the early universe through the Kibble mechanism \cite{kib}.
However unlike other type of topological defects skyrmions are subject
to strict boundary conditions which significantly modify their formation
probability \cite{as}. This must be kept in mind. Common to both heavy 
ion collisions and the electroweak phase transition there is an order
parameter or vacuum expectation value $\lan \f \ran$. Below the
phase transition temperature this takes on a value and a direction
in the normal vacuum. But at temperatures above the phase transition,
there is sufficient energy available so that $\f$ can point in any
directions in the order parameter space. As the system cools down different
spatial regions can have different $\lan \f \ran$'s since they need 
not be correlated. In heavy ion collisions from this point onward
there are two scenarios which can happen and they are not necessarily
mutually exclusive. 

The first scenario is the system would try to restore the vacuum value of
$\lan \f \ran$, $\lan \f \ran = \lan \s \ran$ for the chiral phase transition,
by eliminating spatial domains with anomalous values of $\lan \f \ran$.
This is done through pion radiation. For example if $\f$ is pointing in
the $\p^+$ direction in a domain, this is simply eliminated by emitting
low energy $\p^+$. This has led to the proposal of detecting the chiral
phase transition by detecting the formation of these domains. One examines
the fluctuation in the yield of the three species of pions on a collision
event by collision event basis. This idea was first raised in Ref. \cite{aa,ar}
and gained much momentum through the work of Ref. \cite{bj1,bk}. Much of its
appeal came from cosmic ray data where in a few event there was the absence
of $\p^0$. Such isospin breaking occurrences are known as the Centauro events.
It was thought that one could recreate these Centauro events in heavy ion
collisions or in high energy hadron-hadron collision experiments
\cite{ar,bk,bj1,bkt1,bkt2}. Soon it was shown that in order to detect
these pion radiations amongst the many copiously produced pions and
hadrons via other mechanisms, large domains of fixed chiral orientation
pointing away from the vacuum direction were necessary \cite{rw1,rw2,gm}.
Due to the fact that misalignment of the chiral condensates with the
vacuum condensate is central to this phenomenon, it was coined
Disoriented Chiral Condensates (DCC) soon after the first proposal
\cite{bkt2,rw2}. Motivated by the Centauro events, the primary
observable for DCC is not surprisingly the ratio of the yield of
charged-to-neutral pions. Much work has been done on DCC and especially
analysis on how to extract information from data. Based on 
the fact that the chiral condensates can point in any directions
in the order parameter space which is $S^3$, one expect a distribution
\be \frac{1}{N} \frac{dN}{df} = \frac{1}{2 \sqrt{f}}
\label{eq:dcc-dist}
\ee
for the fraction of neutral-to-total pion $f$ \cite{ar,bk,rw1}.
Unfortunately so far all searches for DCC found no trace of the
existence of these chiral condensates: none in heavy ion
collisions experiments conducted at the Super Proton Synchrotron (SPS) at
CERN \cite{wa98a,wa98b,wa98c} and none at the Tevatron in Fermilab \cite{mmx}.
Instead of the \eref{eq:dcc-dist} these experiments only found a binomial
distribution for $f$ signalling the absence of large chiral domains.

The second scenario is instead of getting rid of the anomalous
order parameter by radiating pions away, the different chiral orientations
of the DCC at the interfaces of the domains may be sufficiently
different that non-trivial topological defects can develop as the
system evolves towards the normal vacuum \cite{ks}. Such defects would be
prevented from collapsing altogether and converting into pion radiations
due to the nonlinearity of the underlying field theory that dictates the
chiral dynamics. These defects would emerge as anomalous production 
of baryons and antibaryons via $B=1$ skyrmion and $B=-1$ antiskyrmion 
through DCC. Although baryon production through skyrmion is not new,
it has only been connected to the much discussed DCC and to 
hyperon data at the SPS recently in Ref. \cite{kw1,kw2,kw3}. The data at 
the SPS may be showing an anomalous baryon-antibaryon production mechanism 
which leads to deviation from the expected rare hyperon yields \cite{kw1,tr}. 

These works use an approach based much on intuition and rough
estimates while evidently a more solid theoretical foundation 
is preferred. For example one does not know exactly what the 
classical skyrmion solutions correspond to in terms of physical
baryons. Therefore one has no concrete way to determine what baryons will 
be produced if skyrmions are formed in heavy ion collisions. 
For instance which baryons exactly among the octet or decuplet baryons
would be produced? If more than one is produced, what is the probability
of a $\Delta$ being produced rather than a $\Sigma$? Why should a proton
be produced but not a $\Xi$? Our attempt to concretize the idea of skyrmion 
production and better connecting it to experimental data has met with a 
serious challenge. 

A survey of the existing literature revealed that there are various 
past attempts at identifying the skyrmion or chiral soliton with a 
superposition of baryon states. In Ref. \cite{ct,gr,bir,fug,ueh}
soliton solutions were found from the linear sigma model and they were
identified with the hedgehog baryons. In these papers each hedgehog is a
three valence quark state surrounded by a coherent pion cloud. The known
baryon states are then projected out of the hedgehog. This method
enabled these authors to study the static properties of the baryons. 
It also gave the probabilities of the distribution of baryons inside 
the chiral soliton. However this method is based on the linear sigma model 
and cannot be used in the Skyrme model because the latter is is essentially 
the non-linear sigma model with additional terms. Another approach is that 
of Ref. \cite{abo,obba} which identifies the skyrmions as $U(4)$ group theoretic 
coherent states of the Perelomov type \cite{ap1,ap2}. The group algebra is 
realized by some auxiliary boson fields and the coherent states are states 
created by these bosons from the vacuum. To study the baryons, one projects 
out the states with the appropriate quantum numbers from these bosonic
coherent states. This method is flexible in that the boson number can be 
equated with $N_c$ the number of colors. Thus one can go to the large 
$N_c$ limit. In spite of some very nice features of these approaches, 
we will adopt another. One that is more closely connected 
to the original Skyrme model and the baryon wavefunctions derived from 
it. The collective quantization of Ref. \cite{anw} produced a set of baryon 
wavefunctions that exist on the compact manifold of $S^3$. As we will
explain in this paper, we are of the opinion that it is better to 
identify a skyrmion directly as a coherent state of baryons. There
will be no auxiliary boson unlike Ref. \cite{abo,obba}, no explicit coherence
in a pion cloud but only one directly of baryon states. To solve this problem 
coherent states of baryons on compact manifolds will be required. 
Very fortunately the long duration between the skyrmion was
first discovered and our arrival at this problem has permitted much
advances in the field of mathematical physics. Because of which it is our
opinion that the time is ripe for tackling this problem using such an approach.

Previously the crucial idea of the classical skyrmion as a coherent
state of baryons and higher resonances was presented in Ref. \cite{me1}. An
outline to the solution was given. The main goal is therefore 
to show how to solve this problem in detail. Since this is evidently of 
interest in its own right, we will concentrate exclusively on obtaining 
the solution. This will be done in two stages. The quantization on the 
compact manifolds will be in this paper. The actual construction
of the coherent states of baryons will be done in a second paper \cite{me2}. 
Applications to heavy ion collisions and other considerations will be 
pursued elsewhere \cite{me3}. 

Because of the indices in space and the abstract group space in 
the paper, it is clearest to fix the notations throughout. We will  
use $\m,\n$ for Lorentz indices, $b,c,d,e =$ 0,1,2,3 for indices in the 
four-dimensional space of $SU(2)$, and $i,j,k,\dots =$ 1,2,3,$\dots$ for 
spatial indices. Repeated indices imply the Einstein summation convention. 

We will first briefly review the Skyrme Model in Sec. \ref{s:sk-md} 
followed by quantization via the collective coordinate approach 
The eigenstates will be obtained which is the first step in making connection 
with the physical states. In the following sections, quantization using both
the $SU(2)$ and $SO(3)$ collective coordinates will be shown. While the
first one has been done before, the $SO(3)$ quantization in all its details
has not appeared in any literature that we know of. Showing them together
here permits a direct comparison. In both cases, the classical form of 
the theory will be shown first, then the Dirac brackets will be constructed 
for a proper quantization of a constrained system. The quantum commutators 
then follow from the classical Dirac brackets. Differential representations 
to the operators will be found next and the eigenstates will be derived from 
the Schr\"odinger equations. Finally in the appendices some technical details 
and wavefunctions are given.

\section{The Skyrme Model}
\label{s:sk-md}

In 1962 Skyrme introduced a Lagrangian density in his attempt to
unify mesons and baryons. The Lagrangian density is
\be {\cal L}_S = \frac{f_\p^2}{4} \; \tr(\del_\m U\del^\m U^\dagger)
                +\frac{1}{32g^2}  \; \tr[U^\dagger \del_\m U,U^\dagger \del_\n U]^2
\label{eq:l_s}
\ee
where
\be U = \exp \{i \mbox{\boldmath $\t$} \cdot \f /f_\p\}
      = ( \s + i \mbox{\boldmath $\t$} \cdot \mbox{\boldmath $\p$})/f_\p
        \; ,
\ee
$f_\p$ is the pion decay constant and $g$ is now known to be the
$\rho$-$\p$-$\p$ coupling. The first term is the usual non-linear sigma
model and the second term was introduced by Skyrme \cite{sk1,sk2}.
It is necessary in order for soliton-like classical solutions to the
Euler-Lagrange equation to exist. To find the solution to the equation of
motion, Skyrme used the so-called Hedgehog ansatz to impose maximal
symmetry to the solution. It has the form
\be U = U_S = \exp \{i \mbox{\boldmath $\t$}\cdot \mbox{\boldmath $\hat r$}\;
                     F(r) \}
\label{eq:u_s}
\ee
where $F(r)$ is a radial function. The energy or mass of the static 
soliton can be worked out by substituting this ansatz into the \eref{eq:l_s}
\bea M &=& -\int d^3 r {\cal L}_S(U=U_S)                        \nonum \\
       &=&  4\p \int^\infty_0 dr
          \Big \{  \half f_\p^2 [r^2 (\mbox{$\frac{dF}{dr}$})^2 + 2\sin^2 F]
                                                           \nonum \\
       & & \hspace{1.5cm}
           +\frac{1}{2g^2} \sin^2 F\; [2 (\mbox{$\frac{dF}{dr}$})^2
           +\mbox{$\frac{\sin^2 F}{r^2}$} ]
          \Big \}   \;.
\eea
In order for the solution to be stable, the energy associated with $U_S$ 
is minimized by a functional variation of $F$ with the fixed 
boundary conditions 
\be F(r \ra \infty) \ra 0  \mbox{\hspace{.8cm} and \hspace{.8cm}}
    F(r=0) = B \p  \;. 
\label{eq:bc}
\ee
This ensures the stability of the solution. The boundary conditions are 
essential for the topological charge to be given by the winding number $B$ 
\cite{sk1,sk2,bha,bmss}. Because the Skyrme model belongs to a class 
of effective Lagrangians that approximate QCD at low energies, 
the solitons or skyrmions of the theory have physical meaning.
In the introduction $B$ has been identified as the baryon
number and a skyrmion is generally considered the ``classical limit'' of
a baryon \cite{sk1,sk2,gw,bnrs1,bnrs2,w1,w2}. Unlike a baryon, a 
skyrmion is a classical object without any quantum number other than $B$.
Therefore a first step towards identifying the classical skyrmion is to 
quantize it.

\subsection{The Hamiltonian} 
\label{ss:qt}

Quantization of the skyrmion amounts to finding the Hamiltonian
and the eigenstates. A static skyrmion has no time dependence and therefore
cannot be quantized. Such a dependence must be introduced. One cannot, however,
have an arbitrary time evolution of $U$ without the risk of losing the 
skyrmion completely or introducing unwanted complications during the time 
evolution (though the topological charge is conserved due to the fact that 
the third homotopy group of $SU(N)$ is the group of integers $\p_3(SU(N))=Z$). 
A solution is to let the time evolution be restricted to a change  
from one skyrmion solution to another. Given that $U_S$ is a solution of 
the Euler-Lagrange equation,  
\be U'_S= A(t) U_S A^\dagger(t)
\ee
is also a solution for any element $A(t)$ of $SU(N)$ because of the form 
of \eref{eq:l_s}. This method of providing a time dependence to the skyrmion 
and yet preserving the essential form of the solution \eref{eq:u_s} was 
first shown in Ref. \cite{anw} and is the well-known collective coordinate 
method of quantization. For simplicity, we will avoid the problems of
three-flavor skyrmions \cite{mnp,wall,ck,nnrv,snnr,vk,hw} and deal
exclusively with only two flavors. 

To quantize the theory, the Hamiltonian is required which can be obtained 
from the Lagrangian. Inserting now $U'_S$ into \eref{eq:l_s} and integrating 
over spatial volume gives the Lagrangian  
\be L = \l \tr (\dot A \dot A^\dag ) - M
\label{eq:sL1}
\ee
where $\l =\frac{8\p}{3 g^3 f_\p} \Lambda$ and
\be \Lambda = \int^\infty_0 d\tilde r\; \tilde r^2 \sin^2 F
              \Big \{1 +(\mbox{$\frac{dF}{d\tilde r}$})^2
                       + \mbox{$\frac{\sin^2 F}{\tilde r^2}$} \Big \}
\ee
with $\tilde r= g f_\p r$. The skyrmion solution $U_S$ is now completely
hidden inside the constants $\l$ and the mass $M$. With only the
kinetic term depending on the collective coordinate $A(t)$, the 
quantization is straightforward. It is more convenient to rewrite the 
element of $SU(2)$ as 
\be A = a_0 + i \bfa \cdot \bft  \;.
\label{eq:A2a}
\ee
Because of the unitary nature of $SU(2)$, the components $a_b$ are 
subject to the constraint
\be c_1 = a_0^2+\bfa^2 = a_b a_b = 1   \;. 
\label{eq:con}
\ee
These coordinates are therefore restricted to the surface of a
four-dimensional unit sphere or $S^3$. In terms of $a_b$'s \eref{eq:sL1} 
becomes
\be L = 2\l \dot a_b \dot a_b - M  \;.
\label{eq:sL2}
\ee
The conjugate momentum to the coordinate $a_b$ is readily found to be
$\p_b = 4 \l \dot a_b$. The Hamiltonian is therefore
\be H = \frac{1}{8\l} \p_b \p_b +M   \;.
\label{eq:ham}
\ee

Ordinarily quantization would be straightforward, one simply imposes 
the commutators 
\be [\hat a_b, \hat a_c]  =0 \;, \hspace{1cm}
    [\hat a_b, \hat \p_c] = i \d_{bc}  \;,
\label{eq:n8-com}
\ee
then finds a differential representation for $\p_b$ and solves the
resulting Schr\"odinger equation for the eigenstates. This would be
essentially the approach of \cite{anw} which we have followed closely
up to now. The presence of the constraint \eref{eq:con} changes 
everything. \eref{eq:n8-com} are the correct commutation relations only 
if $A$ specifies coordinates in Euclidean space $\BR^4$ which is not
the case. Because the space is $S^3$, we will follow the prescription 
of Dirac for quantization in the presence of constraints \cite{di,di2}.

\subsection{The Dirac Brackets} 
\label{ss:db_su2}

To prepare for the quantization, one first find 
all the secondary constraints from the given number of primary
ones. Only after all the constraints have been found can the Dirac brackets 
be constructed from the Poisson brackets. It is the Dirac brackets and not 
the Poisson brackets that will be carried through into the quantum theory as 
commutators upon quantization \cite{di,di2}. In our problem, there is only 
one primary constraint which is \eref{eq:con}. To find the secondary 
constraints, one constructs the total Hamiltonian 
\be H_T = \frac{1}{8\l} \p_b \p_b +M + \rho (a_b a_b-1)   \;. 
\ee
This, in general, is simply the Hamiltonian supplemented with the primary 
constraints via the use of one Lagrange multiplier per constraint. To derive 
the second class constraints, one makes use of the basic Poisson bracket 
\be  \{a_b, \p_c\} = \d_{bc}   \;,
\ee
and the recursion formula
\be  c_{i+1} = \{c_i, H_T\} = 0  \;,
\ee
starting with $i=1$, repeatedly until no more new constraints are produced. 
After dropping non-essential constant factors, we find
\be  c_2 = \shalf (a_b \p_b + \p_b a_b ) = a_b \p_b = \p_b a_b = 0  
\label{eq:c2}
\ee
and
\be  c_3 = \frac{1}{4\l} \p_b \p_b - 2 \rho\, a_b a_b = 0 \;. 
\ee
This process terminates at $c_3$ because $c_2$ is reproduced at $i=4$ 
\be  c_4 = c_2 = 0   \;. 
\ee 
According to Ref. \cite{di,di2} $c_1$ and $c_2$ are the second-class constraints 
necessary for constructing the Dirac brackets while $c_3$ is to provide a solution 
to the Lagrange multiplier $\rho$. To arrive at the Dirac brackets, the 
non-singular anti-symmetric matrix of Poisson brackets between second-class 
constraints must be constructed. This is  
\be m = \bem \{c_1,c_1\} & \{c_1,c_2\}    \\   
             \{c_2,c_1\} & \{c_2,c_2\}    \\    
        \eem        
      = \bem    \;0      &    \;2         \\ 
                 -2      &    \;0         \\   
        \eem                              \;. 
\ee
Equipped with this matrix, the Dirac bracket between any two quantities, 
say $A$ and $B$ is 
\be \{A, B\}_D = \{A, B\} - \{A, c_i\} m^{-1}_{ij} \{c_j, B\}  \;. 
\label{eq:DB-df}
\ee
$m^{-1}$ is the inverse matrix of $m$ and is given by 
\be m^{-1} = \frac{1}{2} \bem    \;0    &   -1    \\
                                 \;1    &   \;0   \\  
                         \eem               \;.   
\ee
Applying this result to all combinations of $a_b$ and $\p_c$, one gets 
\bse
\bea \{ a_b,  a_c\}_D &=& \{a_b,  a_c\} =  0                   \\ 
     \{ a_b, \p_c\}_D &=& \{a_b, \p_c\} - \{a_b, c_2\} m^{-1}_{21} \{c_1, \p_c\} 
                                                        \nonum \\
                      &=& \d_{bc} - a_b a_c                    \\ 
     \{\p_b, \p_c\}_D &=& - \{\p_b, c_1\} m^{-1}_{12} \{c_2, \p_c\} 
                                                        \nonum \\
                      & & - \{\p_b, c_2\} m^{-1}_{21} \{c_1, \p_c\} 
                                                        \nonum \\ 
                      &=& -a_b \p_c + a_c \p_b          \;. 
\eea
\ese
These are the Dirac brackets between the basic positions and conjugate
momenta on $S^3$.

\subsection{Angular Momentum}

On $S^3$ rotation is no different from that in $\BR^4$ and the group is
$SO(4)$ which has six independent generators. These six angular momenta
are direct generalization of the three generators of rotation
in the familiar three-dimensional Euclidean space
\be  L_{bc} = a_b \p_c - a_c \p_b  \;. 
\label{eq:am}
\ee
Using $c_2$, one can verify that 
\be \p_d^2 = \half L_{bc} L_{bc}  
\ee
so that the Hamiltonian can also be written as 
\be H = \frac{1}{16\l} L_{bc} L_{bc} + M  \;. 
\label{eq:ham-L}
\ee
Based on the geometry of the compact space one can expect that the 
angular momentum is more fundamental than the linear ones. This will
be clearly seen later on. This is also the reason why \eref{eq:ham-L} 
is preferred over \eref{eq:ham}.

\subsection{Constrained Quantization}
\label{ss:con-qt-su2}

Quantization in the presence of constraints amounts to making the change 
from classical Dirac brackets to commutators. That is 
\be  \{A, B\}_D  \Lra \frac{1}{i} [\hat A,\hat B ]    \;. 
\ee
The circumflex here over $A$ and $B$ on the right designate them to be 
operators. This reproduces the result for the commutators between 
$\ha_b$ and $\hp_c$ in Ref. \cite{hkp} 
\bse
\label{eq:a-p-com}
\bea [\ha_b, \ha_c] &=& 0                                   \\ {}
     [\ha_b, \hp_c] &=& i (\d_{bc}-\hat a_b \hat a_c)  
\label{eq:a-p-com-2}                                        \\ {} 
     [\hp_b, \hp_c] &=& i (\ha_c \hat \p_b-\ha_b \hat \p_c) 
                     =  i \hL_{cb} 
\eea
\ese
$\hL_{bc}$ are the operators of the angular momenta. These are the 
basic commutators of the $SU(2)$ collective coordinate quantization. 
One can easily find a differential representation that satisfies
the commutators. In fact the representation 
\bse
\bea \ha_b &=&  a_b   \;,        \\ 
     \hp_b &=& -i(\d_{bc}-a_b a_c) \del_c+ i \a a_b 
\label{eq:p-op} 
\eea
\ese 
with any constant $\a$ will satisfy them. The choice of $\a$ and some
related issues will be discussed in the Appendix \ref{a:choice}. This is 
an unfamiliar representation compared to the usual derivative  
form for $\p_b$. However in spite of this, the angular momentum operators 
are still given by 
\be  \hat L_{bc} = \hat a_b \hat \p_c - \hat a_c \hat \p_b
                 = -i (a_b \del_c-a_c \del_b) \;.
\label{eq:am-op} 
\ee

\subsection{Eigenstates}
\label{ss:es}

With the differential representation of $\hL_{bc}$, the Hamiltonian 
in \eref{eq:ham-L} becomes 
\bea \hat H &=& -\frac{1}{8\l} [(\d_{bc}-a_b a_c)\del_b\del_c -3 a_d \del_d]
                +M                                                   \nonum \\
            &=& -\frac{1}{8\l} \nabla^2 + M
\label{eq:ham2}
\eea
where the $\nabla^2$ can be thought of as the Laplacian on $S^3$. Although
its form is not of a Laplacian, it does act like one on $S^3$. The fact that
$\nabla^2$ respects the compact space of $S^3$ permits a problem-free  
(see Appendix \ref{a:qt-s3}) differentiation procedure. As a result,
differentiation can now be performed in exactly the same way as in Euclidean
space. The Schr\"odinger equation can be solved. The wavefunctions are
polynomials in powers of the combinations of products of $(a_b+i a_c)$ 
\cite{anw}. For example with integral power $l$
\be -\nabla^2 (a_b+i a_c)^l = l(l+2) (a_b+ia_c)^l
\label{eq:t1}
\ee
where $b\ne c$ \cite{anw} or with $b\ne c\ne d\ne e$
\bea & & -\nabla^2 (a_b+i a_c)^{l-p} (a_d+i a_e)^p       \nonum \\
     & & \hspace{1.5cm} = l(l+2)  (a_b+i a_c)^{l-p} (a_d+i a_e)^p
\label{eq:t2}
\eea
so the energy only depends on the degree of the polynomial $l$
and not on the which $a_b$'s that make up the polynomial.
With the various combinations of $(a_b+i a_c)$ for a given $l$, there is
a high degree of degeneracy in the energy eigenstates.
To label these states, there must be other quantum numbers beside
energy.

While the momentum operators $\hat \p_b$ acquired unfamiliar form 
in \eref{eq:p-op}, $\hat L_{bc}$ remains the same as those  
in Euclidean space. From their combinations spin $J_i$ and isospin $I_i$
operators can be constructed (see Appendix \ref{a:op}). Using the
corresponding differential operators, it can be easily checked with
\eref{eq:ham2} that
\be -\nabla^2 = 2 (\hat J^2+\hat I^2)
              = 4 \hat J^2  = 4 \hat I^2   \;.
\label{eq:lap-JI}
\ee
The Hamiltonian then becomes
\be  \hat H = \frac{1}{4\l} (\hat J^2+\hat I^2) + M
            = \frac{1}{2\l} \hat J^2 + M
            = \frac{1}{2\l} \hat I^2 + M
\label{eq:ham3}
\ee
and the energy eigenvalues are
\be  E_l = \frac{1}{8\l} l(l+2) + M \;.
\ee
Since
\be  [\hat J^2, \hat J_i] = [\hat I^2, \hat I_i]
   = [\hat J_i, \hat I_j] = 0  \;,
\ee
the Hamiltonian commutes with spin and isospin operators
\be [\hat H, \hat J_i] = [\hat H, \hat I_i]=0
\ee
so eigenstates can be labeled by quantum numbers of energy $l$, the
third component of spin $m$ and isospin $n$ such as $|l,m,n\ran$.
For reason that will become clear below, it is better to use
$|l/2,m,n\ran$ instead.

Acting on a test wavefunction with $\hat J_3$ and $\hat I_3$,
we have
\bea \hat J_3 (a_1 \pm i a_2)^l &=& \pm \mbox{$\frac{l}{2}$} (a_1 \pm ia_2)^l  \\
     \hat I_3 (a_1 \pm i a_2)^l &=& \pm \mbox{$\frac{l}{2}$} (a_1 \pm ia_2)^l
\eea
or
\bea & & \hat J_3 (a_1\pm i a_2)^{l-p} (a_0\pm ia_3)^p         \nonum \\
     & & \hspace{0.7cm} = \pm \mbox{$\frac{l}{2}$}
                          (a_1 \pm i a_2)^{l-p} (a_0\pm ia_3)^p \\
     & & \hat I_3 (a_1\pm i a_2)^{l-p} (a_0\pm ia_3)^p         \nonum \\
     & & \hspace{0.7cm} = \pm (\mbox{$\frac{l}{2}$}-p)
                          (a_1 \pm i a_2)^{l-p} (a_0\pm ia_3)^p  \;.
\eea
The degree of the polynomial $l$ not only gives the energy quantum
number $l$ but also the highest or lowest spin and isospin
$\pm l/2$ of a (iso)spin multiplet. This can also be seen by varying
the integer $p$ between $0$ and $l$. Also from \ethref{eq:t1}{eq:t2}{eq:lap-JI}
we see that the eigenvalues of $\hat J^2$ and $\hat I^2$ for the above
wavefunctions are identical and equal to $j(j+1)=l/2(l/2+1)$ or $j=l/2$.
Other examples of the (iso)spin operators acting on test wavefunctions can be
found in the Appendix \ref{a:wfn}. Even $l$ gives integral and odd $l$ gives
half-integral spin and isospin states. Both bosons and fermions are therefore
admitted in this system which was shown to be possible long ago in Ref. 
\cite{fr}. However not all these states are physical. Baryons are fermions and 
must have half-integral (iso)spin. Therefore only odd degree polynomials are
physical wavefunctions. For example a spin up and spin one-half state has 
$l=1$ or $j=m=n=1/2$ and is represented by
\be \lan a|{\sst \half,\half,\half}\ran = \frac{1}{\p} (a_1+i a_2)
\ee
and a state with spin $-1/2$ and isospin $3/2$, $l=3$ or $j=n=3/2$ and 
$m=-1/2$ has the wavefunction 
\be \lan a|{\sst \frac{3}{2}},{\sst \frac{3}{2}}, {\sst -\half} \ran
     = -\frac{\rtsx}{\p} (a_1 + i a_2) (a_0 - i a_3)^2
\ee 
where the Dirac ket notation $|a\ran$ is used to represent the position
vector on $S^3$. Other spin-half wavefunctions can be written down.
Clearly $(a_b\pm i a_c)^l$ is the highest or lowest (iso)spin state
of a multiplet. One can reproduce the wavefunctions of the whole multiplet
for a given degree $l$ by using the (iso)spin raising or lowering operators
as usual. In Appendix \ref{a:wfn} the complete set of normalized spin one-half
and three-half wavefunctions are given. Quantization therefore permits the 
physical states to be found as the eigenstates of the Skyrme Hamiltonian.

\section{Skyrme Model with $SO(3)$ Collective Coordinates} 
\label{s:so3}  

\subsection{From $SU(2)$ to $SO(3)$}
\label{ss:su22so3}

The general form of the skyrmion solution introduced in Sec. \ref{s:sk-md}
in \eref{eq:u_s} is invariant under a simultaneous $SU(2)$ and spatial rotation.
Therefore applying a $SU(2)$ rotation alone must generate a rotation in space.
This is equivalent to mapping from $SU(2)$ to $SO(3)$. Using the trigonometric
form of \eref{eq:u_s}
\be U_S = \mathbbm{1} \cos F(r) +i \mbox{\boldmath $\t$} \cdot
          \mbox{\boldmath $\hat r$} \sin F(r)  \;
\ee
the action of a $SU(2)$ rotation by an element $A$ is
\be A\, U A^\dag = \mathbbm{1} \cos F(r) +i (A\, \mbox{\boldmath $\t$} \cdot
          \mbox{\boldmath $\hat r$} A^\dagger) \sin F(r) \;.
\ee
$\mathbbm{1}$ denotes generally the identity matrix of a group throughout
this paper. Its dimension depends on the context. Here it
is the $2\times 2$ identity matrix of $SU(2)$. The second
term can be further decomposed in terms of the Pauli spin matrices as
\be  A \t_i A^\dag = \t_j R_{ji}(A)  \;.
\label{eq:su22so3}
\ee
The coefficients of this decomposition $R_{ij}$ which is a function of
$A$ form a $3\times 3$ matrix which can be considered as an element of the
$SO(3)$ group. Imagine now that the $A$ are the collective coordinates for
quantizing the skyrmion. Then the map \eref{eq:su22so3} will permit us to
rewrite everything in terms of the rotation matrix $R$. The collective
coordinates will now have nine components $R_{ij}$ (not all independent)
instead of the four $a_b$ (also not all independent). This method of mapping
from $SU(2)$ to $SO(3)$ is known in the literature \cite{obba,bmss}. In the 
previous section we have already seen how the $SU(2)$ Skyrme theory allows only 
coherent states formed of both boson and fermion states. These states
cannot be the equivalence of a classical skyrmion which must only consist
of physical baryon states. Let us now construct instead the $SO(3)$
equivalence of the classical Skyrme Lagrangian and Hamiltonian.
Then quantization will give us operators and eigenstates in $SO(3)$.

\subsection{The Classical Theory}
\label{ss:so3-cth}

To derive the $SO(3)$ equivalence of \eref{eq:sL1}, we first rewrite
this Lagrangian using the unitary property of $A$ as
\be L = -\l \tr(A^\dagger \dot A)^2 - M   \;.
\ee
Next we use the identity \cite{bmss}
\be  A^\dag \dot A = \shalf \t_i\; \tr (\t_i A^\dag \dot A) \;,
\label{eq:AdA-id}
\ee
which can be easily verified so that we have
\be L = -\shalf \l \Big (\tr (\t_i A^\dag \dot A)\Big )^2 - M   \;.
\label{eq:sL3}
\ee
To proceed further differentiating both sides of the map \eref{eq:su22so3}
with time, followed by multiplying $A^\dagger$ from the left and $A$ from
the right and using \eref{eq:su22so3} again to get
\be   A^\dagger \dot A \t_i + \t_i \dot A^\dagger A
    = \t_l R_{lj}(A^\dagger) \dot R_{ji}(A)          \;.
\ee
Taking the trace after multiplying on both sides with $\t_k$, one arrives at
\be  R_{kj}(A^\dag) {\dot R}_{ji}(A) = -i\e_{kij} \tr(\t_j A^\dag \dot A) \;.
\ee
The first term in \eref{eq:sL3} can therefore be replaced to give us
a Lagrangian in terms of the rotation matrix $R$
\be L = -\mbox{$\4$} \l\; \tr (R^{-1}\dot R)^2 -M  \; .
\label{eq:sL4}
\ee
We have use the fact that $R_{ij}(A^\dagger) = R_{ij}^{-1}(A)$ here.
The nine elements of $R_{ij}$ become the (collective) coordinates.
Similar to the $a_b$ these are not independent but are subject
to the constraints of $SO(3)$
\bse
\label{eq:con-so3}
\bea  \det R      &=& 1                         \\
      R R^{-1}    &=& R^{-1} R = {\mathbbm{1}}  \\
      R^{-1}      &=& R^T                  \;.
\eea
\ese
The superscript $T$ in the last line stands for transpose.
Because of the last constraint one can get rid of $R^{-1}$ everywhere.
More explicitly in component form these constraints are
\bse
\label{eq:con-so3-2}
\bea  \e_{ijk} R_{1i} R_{2j} R_{3k} &=& \e_{ijk} R_{i1} R_{j2} R_{k3} = 1        \\
      \e_{ijk} R_{li} R_{mj} R_{nk} &=& \e_{ijk} R_{il} R_{jm} R_{kn} = \e_{lmn} \\
      R_{ik} R_{jk} &=&  R_{ki} R_{kj} = \d_{ij} \;.
\eea
\ese
The first and second constraints are equivalent but the second is useful
for many algebraic manipulations.

From the above constraints we have the familiar 
\be  R_{ik}^{-1} \dot R_{kj} = R_{ki} \dot R_{kj} = - \dot R_{ki} R_{kj} \;. 
\ee
Therefore the Lagrangian can be rewritten as
\be  L = \mbox{$\4$} \l\; \dot R_{ij} \dot R_{ij} -M  \; .  
\ee 
The conjugate momenta to $R_{ij}$ are 
\be  \Pi_{ij} =  \mbox{$\2$} \l \dot R_{ij}         \; .  
\label{eq:Pi-so3}
\ee
The Hamiltonian is readily derived to be 
\be H =  \mbox{$\4$} \l\; \dot R_{ij} \dot R_{ij}  +M   
      =  \mbox{$\frac{1}{\l}$}\; \Pi_{ij} \Pi_{ij} +M   \;. 
\label{eq:ham-so3} 
\ee

\subsection{The Dirac Brackets}
\label{ss:db_so3} 

As a first step toward quantization, we construct the Dirac brackets. 
To do this, we construct the total Hamiltonian in the same way as before 
from \eref{eq:ham-so3} via the Lagrange multiplier method. Let us see how 
many constraints there are. The first one in \eref{eq:con-so3-2} appears to 
contain $2\times 3\times 3\times 3 =54$ of them but in fact there is only 
one because of antisymmetry. The first constraint is therefore 
\be c_1 = \e_{ijk} R_{1i} R_{2j} R_{3k} - 1 = 0 \;. 
\ee
The second constraint apparently contains eighteen but because of the 
symmetry and equivalence of the transposed matrix, there are really only six
\be c_n = R_{ik} R_{jk} - \d_{ij} = 0  
\label{eq:c2-7} 
\ee
where $n=2,3,\dots,7$ for $(ij)=(11),(12),(22),(13),(23)$ and $(33)$,
respectively. That makes seven constraints for nine $R_{ij}$ components.
But $SO(3)$ is a three-parameter compact Lie group so there can only be six 
constraints of the type of $c_n$ for $n=1$--$7$ (constraints that involve  
only $R_{ij}$ and no $\Pi_{ij}$). After dropping $c_7$ we have the total 
Hamiltonian 
\be  H_T = \mbox{$\frac{1}{\l}$}\; \Pi_{ij} \Pi_{ij} +M 
          +\sum_{n=1}^6 \rho_n c_n  \;. 
\ee  
Remember from the $SU(2)$ theory previously that these are not the
only constraints. The total number of constraints can only be determined
by repeated application of 
\be  c_{i+6} = \{ c_i, H_T \} = 0  
\ee
for $i \ge 1$. This will produce many constraints and we are likely to
have to calculate the inverse of a very large matrix of Poisson brackets
in order to construct the Dirac brackets. This step is very arduous if
not impossible. We shall try another strategy.

\subsubsection{The Special Case of $SO(2)$: The Constraints} 
\label{sss:so2-constr}

To keep the calculation within manageable size, we calculate the $SO(2)$ 
Dirac brackets and then extrapolate the results to $SO(3)$ at the end. 
The primary constraints in this case are 
\bea c_1 &=& \e_{ij} R_{1i} R_{2j} - 1 = 0 \;,        \\
     c_n &=& R_{ik} R_{jk} -\d_{ij} = 0    \;,       
\eea
for $n=2,3,4$ for $(ij)=(11),(12),(22)$ respectively. Here apparently
we have four unknowns and four constraint equations so that each $R_{ij}$
is fixed and can be calculated. However $SO(2)$ is a one parameter Lie group. 
One of the constraints must be dependent of the other three and so can
be dropped.  

There is a choice here of what form of the constraints should be used. 
One could drop one of them, say $c_4$,  but keeping the first three
to ensure that there will not be any redundancy, or as is well known 
there are some simple relations between the matrix elements of $SO(2)$
which are equivalent to the above constraint equations. 
Let us try the following constraints which are complete but simpler
\bea c'_1 &=& \e_{ij} R_{1i} R_{2j} - 1 = 0   \;,        \\ 
     c'_2 &=& R_{11} - R_{22}           = 0   \;,        \\
     c'_3 &=& R_{12} + R_{21}           = 0   \;.   
\eea
These evidently imply 
\be  R_{11}^2 + R_{12}^2 = R_{22}^2 + R_{21}^2 = 1   
\label{eq:r2}
\ee
which shows that they are equivalent to the previous version of $c'_n$'s. 

The total Hamiltonian can be written as 
\bea H_T &=&  \mbox{$\frac{1}{\l}$}\; \Pi_{ij} \Pi_{ij} +M 
             +\rho_1 (\e_{ij} R_{1i} R_{2j} - 1)                \nonum \\
         & & +\rho_2 (R_{11}-R_{22}) +\rho_3 (R_{12}+R_{21})   \;. 
\eea 
Now using the basic Poisson brackets 
\be \{R_{ij}, \Pi_{kl} \} = \d_{ik} \d_{jl}   \;.
\ee
(all other brackets vanish) and after dropping the factor $\rho_j/\l$, 
repeated application of  
\be c'_{i+3} = \{c'_i, H_T\}  = 0 
\ee
gives 
\bea c'_4 &=& \shalf \e_{ij} ( \Pi_{1i} R_{2j} + R_{2j} \Pi_{1i}       \nonum \\  
          & & \hspace{.7cm}    +R_{1i} \Pi_{2j} + \Pi_{2j} R_{1i} ) = 0 \nonum \\
          &=& \e_{ij} ( \Pi_{1i} R_{2j} +R_{1i} \Pi_{2j} ) = 0     \;.
\eea
The simpler $c'_2$ and $c'_3$ give 
\be  c'_5  = \Pi_{11} - \Pi_{22} = 0 
\ee
and 
\be  c'_6  = \Pi_{12} + \Pi_{21} = 0    \;. 
\ee
Continuing with $i\ge 4$, we get 
\be  c'_7  = \frac{4}{\l} \e_{ij} \Pi_{1i} \Pi_{2j} -2 \rho_1 
           -\rho_2 (R_{22}-R_{11}) +\rho_3 (R_{21}+R_{12})  = 0 
\ee
where \eref{eq:r2} has been used. According to Dirac, one is permitted to 
make use of the constraints only after the Poisson brackets have been 
evaluated \cite{di}. Doing this now using $c'_2$ and $c'_3$ reduces $c'_7$ to 
\be  c'_7  = \frac{2}{\l} \e_{ij} \Pi_{1i} \Pi_{2j} -\rho_1  = 0 \;. 
\ee 
Next for $c'_5$ and $c'_6$, again we impose the constraints after 
the brackets have been evaluated to get 
\be  c'_8  = \rho_2  = 0  
\ee
and
\be  c'_9  = \rho_3  = 0    \;. 
\ee 
The constraints $c'_7$, $c'_8$ and $c'_9$ provide solution to the Lagrange
multipliers. There are no first class constraints since none of them has
vanishing Poisson bracket with every other constraints. Therefore the 
second class constraints are $c'_n$, $n=1,\dots,6$. We have a $6\times 6$ 
matrix of Poisson brackets of second class constraints $\{c'_i,c'_j\}$.

\subsubsection{The Special Case of $SO(2)$: The Dirac Brackets} 
\label{sss:db_so2}

The Poisson brackets amongst $c'_1$, $c'_2$ and $c'_3$ evidently vanish.
The same is also true between $c'_5$ and $c'_6$. The only potentially
non-vanishing elements are those Poisson brackets involving $c'_4$ and 
the ($c'_2,c'_5)$ and ($c'_3,c'_6$) pairs. Working them out explicitly,  
the matrix of Poisson brackets $m_{ij}=\{c'_i, c'_j\}$ is 
\be  m = \left ( \begin{smallmatrix}
          \phm 0   & \phm 0   & \phm 0   & \phm 2   & \phm 0   & \phm 0  \\
          \phm 0   & \phm 0   & \phm 0   & \phm 0   & \phm 2   & \phm 0  \\
          \phm 0   & \phm 0   & \phm 0   & \phm 0   & \phm 0   & \phm 2  \\
              -2   & \phm 0   & \phm 0   & \phm 0   & \phm 0   & \phm 0  \\
          \phm 0   &     -2   & \phm 0   & \phm 0   & \phm 0   & \phm 0  \\
          \phm 0   & \phm 0   &     -2   & \phm 0   & \phm 0   & \phm 0  \\ 
                 \end{smallmatrix}  \right ) \;.
\ee
This can be easily inverted to give 
\be  m^{-1} = \frac{1}{2} \left ( 
         \begin{smallmatrix}
          \phm 0   & \phm 0   & \phm 0   &     -1   & \phm 0   & \phm 0  \\
          \phm 0   & \phm 0   & \phm 0   & \phm 0   &     -1   & \phm 0  \\
          \phm 0   & \phm 0   & \phm 0   & \phm 0   & \phm 0   &     -1  \\
          \phm 1   & \phm 0   & \phm 0   & \phm 0   & \phm 0   & \phm 0  \\
          \phm 0   & \phm 1   & \phm 0   & \phm 0   & \phm 0   & \phm 0  \\
          \phm 0   & \phm 0   & \phm 1   & \phm 0   & \phm 0   & \phm 0  \\ 
         \end{smallmatrix} \right )  \;. 
\ee 

Trying out a few Dirac brackets using the definition \eref{eq:DB-df} produces
vanishing brackets between the components of the coordinates  
\be  \{R_{ij}, R_{kl}\}_D  = 0  \;. 
\ee 
Between coordinate and momentum pairs, we can have for example 
\bea \{R_{11}, \Pi_{11}\}_D  
     &=&  1 -\{R_{11}, c'_4\} m^{-1}_{41} \{c'_1, \Pi_{11} \}        \nonum \\
     & &\;\;-\{R_{11}, c'_5\} m^{-1}_{52} \{c'_2, \Pi_{11} \}        \nonum \\ 
     &=& \shalf (1 - R_{22} R_{22}) = \shalf (1 - R_{11} R_{11}) \;, \nonum \\
\label{eq:r11p11}
\eea 
\bea \{R_{11}, \Pi_{12} \}_D 
     &=& -\{R_{11}, c'_4\} m^{-1}_{41} \{c'_1, \Pi_{12} \}    \nonum \\
     & & -\{R_{11}, c'_5\} m^{-1}_{52} \{c'_2, \Pi_{12} \}    \nonum \\
     &=& \shalf R_{22} R_{21} = - \shalf R_{11} R_{12}      \;, 
\label{eq:r11p12} 
\eea
\bea \{R_{12}, \Pi_{12} \}_D
     &=&  1 -\{R_{12}, c'_4\} m^{-1}_{41} \{c'_1, \Pi_{12} \}        \nonum \\ 
     & &\;\;-\{R_{12}, c'_6\} m^{-1}_{63} \{c'_3, \Pi_{12} \}        \nonum \\
     &=& \shalf (1 - R_{21} R_{21}) = \shalf (1 -R_{12} R_{12}) \;,\nonum \\ 
\label{eq:r12p12}  
\eea
and also
\bea \{R_{12}, \Pi_{11} \}_D
     &=& -\{R_{12}, c'_4\} m^{-1}_{41} \{c'_1, \Pi_{11} \}  \nonum \\ 
     &=& \shalf R_{21} R_{22} = -\shalf R_{12} R_{11}  \;.  
\eea 
The same must hold if the indices have been interchanged 
$(1,2) \ra (2,1)$ everywhere. One could now attempt to deduce the general 
expression for the Dirac brackets between $R_{ij}$ and $\Pi_{kl}$  
\be  \{R_{ij}, \Pi_{kl}\}_D  = \d_{ik} \d_{jl} 
                              -\shalf (\d_{ik} \d_{jl} +R_{ij} R_{kl} )
                              + \dots \;. 
\label{eq:guess} 
\ee
The first term is from the first Poisson bracket in the definition of the 
Dirac bracket \eref{eq:DB-df} and the third term seems to hold everywhere. 
The second term is there so that this also agrees with 
\etwref{eq:r11p11}{eq:r12p12}. Lastly there might be other terms hence the
ellipses. 

Other Dirac brackets between $R_{ij}$ and $\Pi_{kl}$ can be deduce using
the constraints. From \eref{eq:r11p11} 
\be  \{R_{11}, \Pi_{22}\}_D  = \{R_{22}, \Pi_{11}\}_D  
                             = \shalf (1 - R_{11} R_{22})  \;,
\label{eq:r11p22}
\ee 
the first term of which deviates from \eref{eq:guess}. Then  
from \eref{eq:r11p12}
\bea   \{R_{22}, \Pi_{12} \}_D &=& -\{R_{11}, \Pi_{21} \}_D
    =  -\{R_{22}, \Pi_{21} \}_D                               \nonum \\  
   &=& -\shalf R_{22} R_{12} =  \shalf R_{11} R_{21}         
    =   \shalf R_{22} R_{21}    \;.                           \nonum \\   
\eea 
These are all consistent with \eref{eq:guess}. Next there is also
\eref{eq:r12p12} which can be combined with the constraints to give
\be  \{R_{12}, \Pi_{21}\}_D = \{R_{21}, \Pi_{12}\}_D 
                            = - \shalf (1 +R_{12} R_{21})  
\ee
and
\be  \{R_{21}, \Pi_{21}\}_D = \shalf (1 -R_{21} R_{21})  \;. 
\ee 
Of these equations the second is consistent with our guess in \eref{eq:guess} 
but not the first. This seems to hint at a term of $-\d_{il} \d_{jk}/2$ when 
$i\ne k$ or $j\ne l$ otherwise this would contradict \eref{eq:r11p11}. 
Then \eref{eq:r11p22} suggests another term of the form $\d_{ij} \d_{kl}/2$
under the same condition of $i\ne k$ or $j\ne l$. The sum of
these two terms would ensure all conditions are met. This requires 
extending \eref{eq:guess} to  
\be  \{R_{ij}, \Pi_{kl}\}_D  = \d_{ik} \d_{jl} 
    -\shalf (\d_{ik} \d_{jl} -\d_{ij} \d_{kl} +\d_{il} \d_{jk}  
    +R_{ij} R_{kl} )  \;. 
\ee
This is consistent with all the above Dirac brackets. 
Simplifying as much as possible finally gives 
\be  \{R_{ij}, \Pi_{kl}\}_D  = \shalf (\d_{ik} \d_{jl} 
                              +\e_{ik} \e_{jl} -R_{ij} R_{kl} )  \;. 
\label{eq:RP-DB-so2}
\ee
We have made use of the identity 
\be  \e_{ik} \e_{jl} = \d_{ij} \d_{kl} -\d_{il} \d_{jk}  
\ee
which is only a special case of the three dimensional expression for 
$\e_{ij3} \e_{mn3}$. 

There still remain the Dirac brackets between momenta to be worked out 
\bea \{\Pi_{ij}, \Pi_{kl} \}_D 
     &=& - \{\Pi_{ij}, c'_1\} m^{-1}_{14} \{c'_4, \Pi_{kl}\}    \nonum \\
     & & - \{\Pi_{ij}, c'_4\} m^{-1}_{41} \{c'_1, \Pi_{kl}\}    \;.
\eea 
For example between equal-index pairs, all Dirac brackets vanish 
\be   \{\Pi_{11}, \Pi_{11}\}_D = \{\Pi_{22}, \Pi_{22}\}_D 
    = \{\Pi_{11}, \Pi_{22}\}_D = 0   \;. 
\ee
Between equal pairs, they also vanish 
\be \{\Pi_{12}, \Pi_{12}\}_D =-\{\Pi_{12}, \Pi_{21}\}_D = 0 \;.   
\ee
Next between unequal-index pairs 
\bea \{\Pi_{11}, \Pi_{12}\}_D &=& \shalf (R_{22}\Pi_{21} -R_{21}\Pi_{22}) \nonum \\
     &=& -\shalf (R_{11}\Pi_{12} -R_{12}\Pi_{11})  \;. 
\label{eq:Pi11Pi12}
\eea 
To arrive at the last expression, one uses $c'_2$ and $c'_3$. This last line
seems to be more closely related to the LHS. From here the rest of the 
brackets can be deduced
\bea \{\Pi_{11}, \Pi_{12}\}_D &=& \{\Pi_{22}, \Pi_{12}\}_D   
      =  -\{\Pi_{11}, \Pi_{21}\}_D                       \nonum \\  
     &=& -\{\Pi_{22}, \Pi_{21}\}_D \ne 0  \;. 
\eea

Aiming now for a general expression, from these examples we can 
infer that the Dirac brackets between conjugate momenta are 
\be \{\Pi_{ij}, \Pi_{kl}\}_D = -\shalf (R_{ij} \Pi_{kl}-R_{kl} \Pi_{ij}) \;. 
\label{eq:PiPi-DB-so2} 
\ee
This is satisfied by the Dirac brackets above. This can be further verified 
using \eref{eq:RP-DB-so2} and the Jacobi identity. The latter is one
of the basic requirements for the Dirac brackets \cite{di}.
The Dirac bracket between conjugate momenta are in general 
non-vanishing. This is different from the familiar examples when the 
different components of the momenta are independent.

\subsubsection{Extrapolating The Dirac Brackets from $SO(2)$ to $SO(3)$} 
\label{sss:x-so2-so3}

Returning now to $SO(3)$, we still have the original primary constraints 
of $c_i$, $i=1$--$6$ which ultimately dictate what the Dirac brackets will 
be. We start by considering the simplest ones --- those involving only 
components of $R$. The vanishing of the Dirac brackets between them 
\be  \{R_{ij}, R_{kl}\}_D  = 0  
\ee 
is expected to continue to hold for general $SO(N)$. This can be understood 
as follows. Since the only non-vanishing Poisson brackets between $R_{ij}$ 
and the constraints are those $c_n$ which involve $\Pi_{ij}$, these must 
all be secondary constraints derived from the $c_i$, $i\le 6$ via 
$\{c_i,H_T\}$. In this way, each primary constraint will lead to a secondary 
constraint $c_n$, $6<n<13$ such as 
\be c_7 = \e_{ijk} ( \Pi_{1i} R_{2j} R_{3k} +R_{1i} \Pi_{2j} R_{3k} 
                    +R_{1i} R_{2j} \Pi_{3k} ) = 0 
\label{eq:c7} 
\ee 
and 
\be c_n = \Pi_{ik} R_{jk} +R_{ik} \Pi_{jk} = 0   
\label{eq:c8-12}
\ee 
for $8\le n \le 12$ with $(i,j)$ = (11), (12), (13), (22), (23).  
The next level of constraints $c_n$ with $n>12$ generated from these will 
therefore be equations involving the Lagrange multiplier $\rho_i$, $i\le 6$ 
since the presence of $\Pi$ will bring the multipliers into the equations. 
We expect no more new constraints after that. The matrix of Poisson brackets 
between second class constraints $m_{ij} = \{c_i, c_j\}$ will be 
a $12\times 12$ square matrix. It must have a $6\times 6$ block of zeros 
in the upper left-hand corner between row one to six and column one to six 
because primary constraints have mutually vanishing Poisson brackets. 
After taking the inverse, this block will reappear on the lower right-hand corner 
of $m^{-1}$. All cofactors in this $6\times 6$ corner vanishes because of the 
block of zeros in $m$. But the only way to have non-vanishing Dirac brackets 
between components of $R$ is if the entries in this block are not all zeros 
because $\{R_{ij},c_p\} \ne 0$ only if $7\ge p \ge 12$. Thus the $R$  
components must commute with each other. 

Let us now extrapolate \eref{eq:RP-DB-so2}. The only part of this equation 
that cannot readily be extended to $SO(3)$ is the second term with the 
antisymmetric tensors. In $SO(3)$ the structure constants are the $\e_{ijk}$ 
instead of $\e_{ij}$. The latter is implicitly equal to $\e_{ij3}$ when 
$i,j$ are restricted only to $1$ and $2$. The $\e_{ik} \e_{jl}$ in 
\eref{eq:RP-DB-so2} must be replaced by $\e_{ik3} \e_{jl3}$. But 
this would leave us with the two 3 indices when 
there is no reason for 3 to be any more special than either 1 and 2. 
If one remembers that $SO(2)$ is a subgroup of $SO(3)$ which is equal to
the former when the constraint $R_{33}=1$ is imposed. With this piece
of information, one can now remove the 3 indices by first identifying
$\e_{ik3} \e_{jl3}$ with $\e_{ik3} \e_{jl3} R_{33}$. Then one can 
replace this with $\e_{ikm} \e_{jln} R_{mn}$ when the first pair of indices 
are no longer restricted to 1 and 2. The Dirac brackets become 
\be  \{R_{ij}, \Pi_{kl}\}_D  = \shalf (\d_{ik} \d_{jl} 
                              +\e_{ikm} \e_{jln} R_{mn}-R_{ij} R_{kl} )  \;. 
\label{eq:RP-DB}
\ee
This must hold because one could imagine using the indices, say 2 and 3, 
for $SO(2)$ instead of 1 and 2 then this would reduce to the correct
form even for this artificial case when $R_{11}=1$ and 
$R_{1i}=R_{i1}=0$ for $i=2,3$.  

Lastly the brackets \eref{eq:PiPi-DB-so2} seem to be ready for 
generalization to $SO(3)$ by letting the indices running from 1 to 3. 
Are these the correct Dirac brackets for $SO(3)$?
In fact for $SO(2)$, the form of \eref{eq:PiPi-DB-so2} is not unique.
In this special case, one could equally rewrite it after some
experimentation with \eref{eq:Pi11Pi12} using $c'_2$ and $c'_3$ as
\bea \{\Pi_{ij}, \Pi_{kl}\}_D  
    &=&  \mbox{$\4$} \d_{ik} ( R_{ml}\Pi_{mj} -R_{mj}\Pi_{ml} )   \nonum \\ 
    & & +\mbox{$\4$} \d_{jl} ( R_{km}\Pi_{im} -R_{im}\Pi_{ki} )   \;.  
\label{eq:PiPi-DB} 
\eea
Which of these are the correct Dirac brackets? It can be easily checked
by using the Jacobi identity that \eref{eq:PiPi-DB} are the correct ones.

\subsection{Spin and Isospin Operators}

The Lagrangian \eref{eq:sL4} manifests symmetry under the left
and right $SO(3)$ rotation. An infinitesimal transformation with
parameter $\h_k$, where $k$ index labels the direction of rotation, gives
\bea  R_{ij} &\ra & R_{ij} + \h_k \d^L_k R_{ij} = R_{ij} +\h_k \e_{kil} R_{lj}  \\
      R_{ij} &\ra & R_{ij} + \h_k \d^R_k R_{ij} = R_{ij} +\h_k R_{il} \e_{klj}  \;.
\eea
Because $R$ has only time dependence, these generate the conserved
left and right Noether charges but no three-current
\bse
\label{eq:Q-def}
\bea  Q^L_i &=& \frac{\del L}{\del \dot R_{jk}} \d^L_i R_{jk}
             =  \Pi_{jk} \e_{ijl} R_{lk} = \e_{ijk} R_{kl} \Pi_{jl} \nonum \\
                                                                           \\
      Q^R_i &=& \frac{\del L}{\del \dot R_{jk}} \d^R_i R_{jk}
             =  \Pi_{jk} R_{jl} \e_{ilk} = \e_{ijk} \Pi_{lk} R_{lj}  \;.
                                                                    \nonum \\
\eea
\ese
Time evolution of the primary constraints in \eref{eq:con-so3-2} leads to
a set of secondary constraints \cite{di,di2}. For example the second equation
in \eref{eq:con-so3-2} gives
\be  R_{ik} \Pi_{jk} +\Pi_{ik} R_{jk} = R_{ki} \Pi_{kj} +\Pi_{ki} R_{kj} = 0 \;.
\ee
This together with the primary constraints allow us to derive
\be  Q^L_i Q^L_i = Q^R_i Q^R_i = 2 \Pi_{ij} \Pi_{ij}    \;.
\ee
The Hamiltonian acquires yet other equivalent forms  
\be H = \mbox{$\frac{1}{2\l}$}\; Q^L_i Q^L_i +M
      = \mbox{$\frac{1}{2\l}$}\; Q^R_i Q^R_i +M    \;.
\ee
The significants of the conserved charges will become clear in the
next section.

\subsection{The Connection between the $SU(2)$ and $SO(3)$
collective variables}
\label{ss:connect}

To get an idea of the physics of the conserved charges, we now derive
the connections between the $SU(2)$ conjugate pairs ($a_b,\p_b$) and
those pairs ($R_{ij},\Pi_{ij}$) of $SO(3)$. From the map \eref{eq:su22so3},
one can deduce that
\be R_{ij} = \shalf \tr (\t_i A \t_j A^\dagger )  \;.
\ee
To make the connection more explicit, one can use the component form of $A$
in \eref{eq:A2a} to get the expression \cite{me1}
\be R_{ij} = 2 a_i a_j +\d_{ij} (2 a_0^2-1) -2 \e_{ijk} a_0 a_k \;.
\ee
This expression together with \eref{eq:Pi-so3} give us the conjugate
momentum $\Pi_{ij}$ in terms of $a_b$ and $\p_b$
\bea \Pi_{ij} &=& \mbox{$\4$} \{ a_i \p_j+a_j \p_i
                                +\d_{ij} (a_0 \p_0 +\p_0 a_0)      \nonum \\
              & & \hspace{0.7cm}  -\e_{ijk} (a_0 \p_k+\p_0 a_k) \}  \;.
\eea
Substituting these functions into \eref{eq:Q-def} and after some algebra,
we have
\bea Q^L_i &=& \shalf ( a_i \p_0 -a_0 \p_i - \e_{ijk} a_j \p_k )   \\
     Q^R_i &=& \shalf ( a_i \p_0 -a_0 \p_i + \e_{ijk} a_j \p_k )   \;.
\eea
Comparing the expressions for $J_i$ and $I_i$ in \etwref{eq:J}{eq:I}
(temporarily replacing $\p_b$ with $-i\del_b$ to see this) reveals that
\be  Q^L_i = -J_i \;,\;\;\;\;  Q^R_i = I_i   \;.
\ee
The conserved charges are actually spin and isospin. Note that all
variables here are classical and commuting so that there is no ambiguity
as to the ordering. For this reason, we will refrain from carrying the
connections here to the quantum level when one does not know which
operator ordering is the correct one. However one still identifies
the charge operators with spin and isospin operators as shown here. 
Therefore the Hamiltonian can be written in terms of spin and isospin
as in the $SU(2)$ case
\be H = \frac{1}{2\l} J^2 + M = \frac{1}{2\l} I^2 + M   \;. 
\label{eq:ham-so3-JI}
\ee

\subsection{Constrained Quantization}
\label{ss:con-qt-so3}

Quantization is accomplished by simply replacing the Dirac brackets 
\cite{di2} by $-i$ times the commutators. Therefore 
\bse
\label{eq:com-so3} 
\bea [\hR_{ij}, \hR_{kl}] &=& 0                                           \\ {} 
     [\hR_{ij}, \hP_{kl}] &=& \shalf i (\d_{ik} \d_{jl} 
                             +\e_{ikm} \e_{jln} \hR_{mn}-\hR_{ij} \hR_{kl} )  
                                                                \nonum \\ \\ {}
     [\hP_{ij}, \hP_{kl}]  
    &=&  \mbox{$\4$} i\d_{ik} ( \hR_{ml}\hP_{mj} -\hR_{mj}\hP_{ml} )   \nonum \\ {} 
    & & +\mbox{$\4$} i\d_{jl} ( \hR_{km}\hP_{im} -\hR_{im}\hP_{ki} )   \;.  
\eea
\ese

A simple multiplication and differential representation of the operators are 
\bea \hR_{ij} &=& R_{ij}                                            \\ {}
     \hP_{ij} &=& -\shalf i (\d_{ik} \d_{jl} +\e_{ikm}\e_{jln} R_{mn}
                             -R_{ij} R_{kl} ) \del_{kl} +i \a R_{ij} 
                                                             \nonum \\ {} 
\eea
where the shorthand $\del_{ij} = \del/\del R_{ij}$ is used. 
Promoting \eref{eq:Q-def} first to operators, then after some algebra 
using the identities 
\bse
\bea  \e_{mja} \e _{nlb} \hR_{kl} \hR_{ab}
    &=& \hR_{mn} \d_{jk} - \hR_{jn} \d_{mk}       \\ 
      \e_{mla} \e_{nkb} \hR_{lj} \hR_{ab}
    &=& \hR_{mn} \d_{jk} - \hR_{mk} \d_{jn}       
\eea
\ese 
and the above representation, it can be shown that
\bea \hJ_i &=& \shalf \e_{ijk} ( \hR_{jl}\hP_{kl} -\hR_{kl}\hP_{jl} ) \nonum \\ 
           &=&-\shalf i\e_{ijk} (R_{jl} \del_{kl} - R_{kl} \del_{jl} )  \\ {}
     \hI_i &=& \shalf \e_{ijk} ( \hR_{lj}\hP_{lk} -\hR_{lk}\hP_{lj} ) \nonum \\
           &=&-\shalf i\e_{ijk} (R_{lj} \del_{lk} - R_{lk} \del_{lj} )  \;.
\label{eq:JI-df}
\eea
One can easily check that they satisfy all the commutation relations
expected of the spin and isospin operators (these are of course the same
as those for $SU(2)$ see Appendix \ref{a:op}).

\subsection{Eigenstates}
\label{ss:es-so3}

From \eref{eq:ham-so3-JI} and the $\hJ$ and $\hI$ operators, the 
differential form of the Hamiltonian is 
\bea \hat H &=& -\mbox{$\frac{1}{2\l}$} ( \del_{ij} -2 R_{ij}
                                 -R_{im} R_{kj} \del_{km} ) \del_{ij} + M
                                                                    \nonum \\
            &=& -\mbox{$\frac{1}{2\l}$} \nabla^2 + M     \;.
\eea
To find the wavefunctions, it is more convenient to use specific 
combinations of the components. They are 
\bse
\label{eq:R}
\bea  R^{++}      &=& R_{11}-R_{22} +i(R_{12}+R_{21})  \\
      R^{+-}      &=& R_{11}+R_{22} -i(R_{12}-R_{21})  \\
      R^{-+}      &=& R_{11}+R_{22} +i(R_{12}-R_{21})  \\
      R^{--}      &=& R_{11}-R_{22} -i(R_{12}+R_{21})  \\
      R^{\pm0}    &=& R_{13} \pm i R_{23}              \\
      R^{0\pm}    &=& R_{31} \pm i R_{32}              \\
      R^{00}\;    &=& R_{33}  \;.
\eea
\ese 
Using these, it can be easily verified that  
\be -\nabla^2 (R^{pq})^l = l(l+1) (R^{pq})^l
\ee
where $l \in \BZ$ with the exception of $p=q=0$ when $l> 1$, or
\be -\nabla^2 (R^{\pm q})^{l-r} (R^{\pm 0})^r
    = l(l+1) (R^{\pm q})^{l-r} (R^{\pm 0})^r
\ee
with $r \le l$. The energy eigenvalues are then
\be  E_l = \mbox{$\frac{1}{2\l}$} l(l+1) + M   \;.
\ee
It can thus be established that the wavefunctions are polynomials in 
the products of the components of $R$. They are also eigenstates of the 
third component of spin and isospin 
\bea \hJ_3 (R^{pq})^l &=& p\, l\, (R^{pq})^l       \\
     \hI_3 (R^{pq})^l &=& q\, l\, (R^{pq})^l       \\
     \hJ_3 (R^{\pm q})^{l-r} (R^{\pm 0})^r
                      &=& \pm l (R^{\pm q})^{l-r} (R^{\pm 0})^r     \\
     \hI_3 (R^{\pm q})^{l-r} (R^{\pm 0})^r
                      &=& q\, (l-r) (R^{\pm q})^{l-r} (R^{\pm 0})^r \;.
                                                             \nonum \\
\eea
The polynomial degree $l$ dictates the energy, spin and isospin eigenvalues
just as in the $SU(2)$ theory. This time, however, there are only integer
values of $l$ (functions with half-integer $l$ are not defined). 
The Hilbert space of this system consists purely of bosonic states
$|j,m,n\ran$ where $j,m,n \in \BZ$. These are not physical states 
expected from the Skyrme model.

\section{Summary and Outlook}

In this paper we have achieved the quantization of the Skyrme model
on the $S^3$ as well as on the $SO(3)$ compact manifolds. These have 
been done in a rigorous manner using the Dirac prescription for a 
constrained system. While the results from the $SU(2)$ collective
coordinate quantization are mostly known, this is not true for the same on
the $SO(3)$ manifold. For example, the commutators \eref{eq:com-so3} in this 
case have not appeared in any literature that we know of. This may be related 
to the fact the eigenstates are not physical and therefore there is little 
incentive to pursuit this second method in any detail. This will turn out
not to be completely true. In the next paper to come and indeed was 
mentioned in \cite{me1}, we will need the $SO(3)$ quantized operators, 
if not the unphysical eigenstates, to construct coherent states of baryons. 
The method of coherent states on a compact manifold is a mixture of methods 
used in the context of functional analyses \cite{ha} and in the studies of
the classical limit of quantum gravity \cite{krp,kr,hm,tt0,tt1,tt2,tt3}. 
All these will be shown in the work to come \cite{me2}.

\section*{Acknowledgments}
The author thanks K. Kowalski for pointing out Ref. \cite{tt2,hm}, 
B.C. Hall for clarifying the annihilation operators, T. Thiemann for
Ref. \cite{tt1,tt3}, R. Amado and R. Bijker for explaining their papers 
\cite{abo,obba}, P. Ellis, U. Heinz and J. Kapusta for comments. 
This work was supported by the U.S. Department of Energy under grant 
no. DE-FG02-01ER41190.

\appendix

\section{Choice of the secondary constraint} 
\label{a:choice}  

Quantizing the theory produces certain ambiguities associated with the
constraints. These ambiguities have to do with the ordering of the classical 
c-numbers turned quantum q-numbers. For example while the primary constraint 
\be \hat c_1 = \ha_b \ha_b - \hat 1 = 0   \;, 
\label{eq:hc1}
\ee
does not pose a problem, the secondary constraint \eref{eq:c2} can be 
expressed as any of the following:  
\be \hat c_2 = \ha_b \hp_b = 0   
\label{eq:hc2a}
\ee
or
\be \hat c_2 = \hp_b \ha_b = 0   
\label{eq:hc2b}
\ee
or 
\be \hat c_2 = \shalf (\ha_b \hp_b +\hp_b \ha_b) = 0   \;. 
\label{eq:hc2c} 
\ee
They are not equivalent because of \eref{eq:a-p-com-2}. Let us not insist 
on any one of them. From \eref{eq:a-p-com-2} they imply 
\bse
\bea \hp_b\,\ha_b &=& -i(d-1)          \;,
\label{eq:q-c2-1}                      \\ {} 
     \ha_b\,\hp_b &=&  i(d-1)          \;, \text{\hspace{1cm} or} 
\label{eq:q-c2-2}                      \\ {}
     \ha_b\,\hp_b &=&  \shalf i(d-1)   \;,\;\;\;
     \hp_b\,\ha_b  =  -\shalf i(d-1)   \;, 
\label{eq:q-c2-3} 
\eea
\ese 
respectively. We have decided to introduce the dimension $d$ during the 
intermediate steps of the calculation which can be set to $d=4$ at any time.  
We have seen in the differential representation of $\p_b$ in \eref{eq:p-op}
that there is a free parameter $\a$. Using this representation and
the above equations, one can see that $\a$ is in fact not really
a free parameter. Its value is actually restricted by how the secondary
constraint is implemented. In the three cases, we have 
\bse
\bea \a &=& 0                 \;, \\
     \a &=& d-1               \;, \\ 
     \a &=& \shalf (d-1)      \;,
\eea
\ese
or equivalently 
\bse
\bea \hat \p_b &=& -i (\d_{bc}-a_b a_c) \del_c   
\label{eq:p-op-1}                             \;, \\
     \hat \p_b &=& -i (\d_{bc}-a_b a_c) \del_c + i(d-1) \ha_b  
\label{eq:p-op-2}                             \;, \\
     \hat \p_b &=& -i (\d_{bc}-a_b a_c) \del_c + \shalf i(d-1) \ha_b 
\label{eq:p-op-3}                             \;, 
\eea
\ese
respectively. Among the three choices, the third one seems more natural 
in the sense that \eref{eq:hc2c} is a direct consequence quantum
mechanically of the time-evolved form of \eref{eq:hc1}. 

In the main text, the readers might have noticed that we always expressed
the classical Hamiltonian in terms of $J^2$ first and then upon quantization
substituting everything with their corresponding operators. This is slightly
different from quantizing with the classical Hamiltonian expressed in terms
of the $\p_b \p_b$. We have been careful to avoid the latter because of the
non-uniqueness of the differential representation of $\hp_b$. On the other
hand the differential representation of $\hJ_i$ is independent on the choice
of $\a$. We feel that the Hamiltonian should be free from such ambiguity
and in any case $\hJ_i$ are more fundamental on $S^3$ than $\p_b$. In favoring
$\hJ_i$ over $\hp_b$, the Hamiltonian operator is  
\be \hat H = \frac{1}{2\l} \hJ^2 +M   
           = \frac{1}{8\l} \hp_b \hp_b + M + \frac{1}{8\l} \a (\a-d+1)  \;. 
\ee 
The last equality is more cumbersome looking. The extra terms are needed
so that different values of $\a$ will give the same Hamiltonian because
the $\hp_b$ will change with $\a$.

\section{Quantization on $S^3$}
\label{a:qt-s3}

Quantization of the Hamiltonian \eref{eq:ham} is problematic 
because the coordinates $A$ lives on $S^3$ which is a curved compact
space. It is tempting to follow the usual procedure of canonical
quantization and use the commutators in \eref{eq:n8-com}
\be [\hat a_b, \hat  a_c]  =0 \;, \hspace{1cm}
    [\hat a_b, \hat \p_c] = i \d_{ij}  \;.
\label{eq:a-n8-com}
\ee
This would give the familiar differential representation for $\hat \p_b$
\be  \hat \p_b = -i \frac{\del}{\del a_b} = -i \del_b \;.
\ee
The problem of following this standard procedure is that in the subsequent
use of $\hat \p_b$, it is always necessary to keep in mind that the derivative
is meant to be on $S^3$. Thus the straightforward rules of differentiation
does not necessarily apply and indeed may be modified. For example
while $\del_b a_c =\d_{bc}$ in $\BR^n$, this is not true on $S^3$ 
which can be very confusing. In fact one must use the strange
result $\del_b a_c = \d_{bc}- a_b a_c$ to get the right answer.
The problem is that the commutators \eref{eq:n8-com} do not respect the
constraint of this system which is $a_b a_b =1$. A correct quantization
procedure must yield
\be  [\hat a_b \hat a_b -\hat 1, \hat O] =
     [\hat a_b \hat a_b        , \hat O] = 0  \;,
\ee
where $\hat O$ is any operator, whereas \eref{eq:n8-com} gives
\be  [\hat a_b \hat a_b, \hat  a_c]  =0 \;, \hspace{1cm}
     [\hat a_b \hat a_b, \hat \p_c]  = 2 i \hat a_c  \;.
\label{eq:n8-ccom}
\ee
The first one of these is fine but not the second. In the presence of 
constraints, one quantizes the Dirac brackets instead of the Poisson 
brackets. In short the Dirac bracket is the Poisson bracket but with 
some subtractions so that every constraint of the system is respected 
by the Dirac bracket. Then upon quantization they are automatically
respected by the commutators. In essence this is not dissimilar to the way
of constructing a vector from a given vector so that the result is 
orthogonal to a third. In the main text we showed how to do this properly.

\section{Angular momentum, spin, isospin operators and their
eigenfunctions}
\label{a:op}

In the text we gave the six generators of rotation in $\BR^4$ or those
of the group $SO(4)$
\be  \hat L_{bc} = \hat a_b \hat \p_c - \hat a_c \hat \p_b
                 = -i (a_b \del_c-a_c \del_b)    \;.
\ee
Rewriting them more conveniently in three-vector form
\bse
\bea  \hat L_i &=& \half \ve_{ijk} \hat L_{jk}      \\
      \hat K_i &=& \hat L_{0i}
\eea
\ese
they obey the following commutators
\bse
\bea [\hat L_i,\hat L_j] &=& i\ve_{ijk} \hat L_k         \\ {}
     [\hat L_i,\hat K_j] &=& i\ve_{ijk} \hat K_k         \\ {}
     [\hat K_i,\hat K_j] &=& i\ve_{ijk} \hat L_k         \;.
\eea
\ese
From them one can construct generators of two separate copies of
$SU(2)$
\bea \hat J_i &=& \half (\hat L_i +\hat K_i)
              = \half i \Big (a_i\del_0-a_0\del_i-\ve_{ijk} a_j\del_k \Big )
\label{eq:J}                                            \\ {}
     \hat I_i &=& \half (\hat L_i -\hat K_i)
              = \half i \Big (a_0\del_i-a_i\del_0-\ve_{ijk} a_j\del_k \Big )
\label{eq:I}                                            \;.
\eea
They are the spin and isospin operators with the familiar
commutators
\bse
\label{eq:JI-com}
\bea [\hat J_i,\hat J_j] &=& i\ve_{ijk} \hat J_k         \\ {}
     [\hat I_i,\hat I_j] &=& i\ve_{ijk} \hat I_k         \\ {}
     [\hat J_i,\hat I_j] &=& 0                           \;.
\eea
\ese
One can form raising and lowering operators for spin and isospin
as usual
\be  \hat J_\pm = \hat J_1\pm i\hat J_2 \;,\;\;\;\;
     \hat I_\pm = \hat I_1\pm i\hat I_2  \;.
\ee
Their commutation relations are
\bse
\bea [  \hat J_+,\hat J_-  ] &=& 2  \hat J_3 \;,\;\;\;\;
     [  \hat J_3,\hat J_\pm]  = \pm \hat J_\pm \;,        \\ {}
     [\;\hat I_+,\hat I_-  ] &=& 2\,\hat I_3 \;,\;\;\;\;
     [\;\hat I_3,\hat I_\pm] = \pm\, \hat I_\pm
\eea
\ese
which follows naturally from \eref{eq:JI-com}. As is well known,
the squared operators can be expressed in terms of them
\bse
\bea  \hJ^2 &=& \hJ_- \hJ_+ +\hJ_3 (\hJ_3+1)
             =  \hJ_+ \hJ_- +\hJ_3 (\hJ_3-1)        \nonum \\ {} && \\ {}
      \hI^2 &=& \hI_- \hI_+ +\hI_3 (\hI_3+1)
             =  \hI_+ \hI_- +\hI_3 (\hI_3-1)    \;. \nonum \\ {} &&
\eea
\ese

\section{Wavefunctions on $S^3$}
\label{a:wfn}

The operators $\hat J_3$ and $\hat I_3$ acting on some general
representative wavefunctions is enough to show the range and relation
of the spin and isospin quantum numbers to the polynomial degree $l$.
Acting on wavefunctions of $(a_b+i a_c)$ raised to some power give
\bea \hat J_3 (a_1 \pm i a_2)^l &=& \pm \mbox{$\frac{l}{2}$} (a_1 \pm ia_2)^l  \\
     \hat J_3 (a_0 \pm i a_3)^l &=& \pm \mbox{$\frac{l}{2}$} (a_0 \pm ia_3)^l  \\
     \hat I_3 (a_1 \pm i a_2)^l &=& \pm \mbox{$\frac{l}{2}$} (a_1 \pm ia_2)^l  \\
     \hat I_3 (a_0 \pm i a_3)^l &=& \mp \mbox{$\frac{l}{2}$} (a_0 \pm ia_3)^l
\eea
and on a product of two different $(a_b+i a_c)$'s give
\bea & & \hat J_3 (a_1\pm i a_2)^{l-p} (a_0\pm ia_3)^p         \nonum \\
     & & \hspace{0.7cm} = \pm \mbox{$\frac{l}{2}$}
                          (a_1 \pm i a_2)^{l-p} (a_0\pm ia_3)^p \\
     & & \hat J_3 (a_1\pm i a_2)^{l-p} (a_0\mp ia_3)^p         \nonum \\
     & & \hspace{0.7cm} = \pm (\mbox{$\frac{l}{2}$}-p)
                          (a_1 \pm i a_2)^{l-p} (a_0\mp ia_3)^p \\
     & & \hat I_3 (a_1\pm i a_2)^{l-p} (a_0\pm ia_3)^p         \nonum \\
     & & \hspace{0.7cm} = \pm (\mbox{$\frac{l}{2}$}-p)
                          (a_1 \pm i a_2)^{l-p} (a_0\pm ia_3)^p \\
     & & \hat I_3 (a_1\pm i a_2)^{l-p} (a_0\mp ia_3)^p         \nonum \\
     & & \hspace{0.7cm} = \pm \mbox{$\frac{l}{2}$}
                          (a_1 \pm i a_2)^{l-p} (a_0\mp ia_3)^p
\eea
where $l \ge p$ and $p \ge 0$. Varying $p$ between $0$ and $l$ shows that
$l/2$ is the largest modulus of the third component of spin and isospin of
the multiplet.

In flavor SU(2) Adkins et al \cite{anw} showed that one could obtain
the normalized wavefunctions on $S^3$. We give the complete 
set for $j=0$, $1/2$, $1$ and $3/2$ wavefunctions. The relative 
phases of these wavefunctions in a multiplet are important. 
For $j=0$ this is simply a constant
\be \lan a|0,0,0 \ran = \frac{1}{\p\sqrt{2}}  \;.
\ee 

For $j=1/2$ 
\bea
  \lan a|{\sst +\half, +\half, +\half} \ran \fxd &=&\fxd +\frac{1}{\p} (a_1 + i a_2) \\
  \lan a|{\sst +\half, +\half, -\half} \ran \fxd &=&\fxd -\frac{i}{\p} (a_0 - i a_3) \\
  \lan a|{\sst +\half, -\half, +\half} \ran \fxd &=&\fxd +\frac{i}{\p} (a_0 + i a_3) \\
  \lan a|{\sst +\half, -\half, -\half} \ran \fxd &=&\fxd -\frac{1}{\p} (a_1 - i a_2) 
\eea

For $j=1$
\bea
  \lan a|1, +1, +1 \ran \fxd &=& \fxd \frac{\sqrt{3}}{\p\rttw} (a_1 + i a_2)^2 \\ 
  \lan a|1, +1,\;\;\,0 \ran \fxd &=& \fxd i \frac{\sqrt{3}}{\p} 
                                 (a_1 + i a_2) (a_0 + i a_3)                   \\ 
  \lan a|1, +1, -1 \ran \fxd &=& \fxd -\frac{\sqrt{3}}{\p\rttw} (a_0 + i a_3)^2 
\eea
\bea 
  \lan a|1,\;\;\,0, +1 \ran \fxd &=& \fxd-i \frac{\sqrt{3}}{\p} 
                                 (a_1 + i a_2) (a_0 - i a_3)                   \\ 
  \lan a|1,\;\;\,0,\;\;\,0 \ran \fxd &=& \fxd \frac{\sqrt{3}}{\p\rttw} 
                           \{ (a_0^2 + a_3^2) -(a_1^2 + a_2^2) \}              \\ 
  \lan a|1,\;\;\,0, -1 \ran \fxd &=& \fxd-i \frac{\sqrt{3}}{\p} 
                                 (a_1 - i a_2) (a_0 + i a_3)            
\eea
\bea 
  \lan a|1, -1, +1 \ran \fxd &=& \fxd -\frac{\sqrt{3}}{\p\rttw} (a_0 - i a_3)^2 \\ 
  \lan a|1, -1,\;\;\,0 \ran \fxd &=& \fxd i \frac{\sqrt{3}}{\p} 
                                 (a_1 - i a_2) (a_0 - i a_3)                    \\ 
  \lan a|1, -1, -1 \ran \fxd &=& \fxd  \frac{\sqrt{3}}{\p\rttw} (a_1 - i a_2)^2 
\eea

For $j=3/2$ 
\bea
  \lan a|{\sst +\thtw, +\thtw, +\thtw} \ran \fxd &=&\fxd  \frac{\rttw}{\p} 
                                            (a_1 + i a_2)^3                  \\
  \lan a|{\sst +\thtw, +\half, +\thtw} \ran \fxd &=&\fxd i\frac{\rtsx}{\p}  
                                            (a_1 + i a_2)^2 (a_0 + i a_3)    \\
  \lan a|{\sst +\thtw, -\half, +\thtw} \ran \fxd &=&\fxd- \frac{\rtsx}{\p} 
                                            (a_1 + i a_2) (a_0 + i a_3)^2    \\
  \lan a|{\sst +\thtw, -\thtw, +\thtw} \ran \fxd &=&\fxd-i\frac{\rttw}{\p} 
                                            (a_0 + i a_3)^3 
\eea

\bea
  \lan a|{\sst +\thtw, +\thtw, +\half} \ran \fxd &=&\fxd-i\frac{\rtsx}{\p} 
                                            (a_1 + i a_2)^2 (a_0 - i a_3)    \\
  \lan a|{\sst +\thtw, +\half, +\half} \ran \fxd &=&\fxd- \frac{\rttw}{\p} 
                                    (a_1 + i a_2) (1 - 3 (a_0^2 + a_3^2))  \nonum \\ \\
  \lan a|{\sst +\thtw, -\half, +\half} \ran \fxd &=&\fxd i\frac{\rttw}{\p} 
                                    (a_0 + i a_3) (1 - 3 (a_1^2 + a_2^2))  \nonum \\ \\
  \lan a|{\sst +\thtw, -\thtw, +\half} \ran \fxd &=&\fxd  \frac{\rtsx}{\p} 
                                    (a_1 - i a_2) (a_0 + i a_3)^2 
\eea

\bea
  \lan a|{\sst +\thtw, +\thtw, -\half} \ran \fxd &=&\fxd- \frac{\rtsx}{\p} 
                                    (a_1 + i a_2) (a_0 - i a_3)^2            \\
  \lan a|{\sst +\thtw, +\half, -\half} \ran \fxd &=&\fxd-i\frac{\rttw}{\p} 
                                    (a_0 - i a_3) (1 - 3 (a_1^2 + a_2^2)) \nonum \\ \\
  \lan a|{\sst +\thtw, -\half, -\half} \ran \fxd &=&\fxd  \frac{\rttw}{\p} 
                                    (a_1 - i a_2) (1 - 3 (a_0^2 + a_3^2))    \\
  \lan a|{\sst +\thtw, -\thtw, -\half} \ran \fxd &=&\fxd i\frac{\rtsx}{\p} 
                                    (a_1 - i a_2)^2 (a_0 + i a_3)
\eea

\bea
  \lan a|{\sst +\thtw, +\thtw, -\thtw} \ran \fxd &=&\fxd i\frac{\rttw}{\p} 
                                     (a_0 - i a_3)^3                     \\
  \lan a|{\sst +\thtw, +\half, -\thtw} \ran \fxd &=&\fxd  \frac{\rtsx}{\p} 
                                     (a_0 - i a_3)^2 (a_1 - i a_2)       \\
  \lan a|{\sst +\thtw, -\half, -\thtw} \ran \fxd &=&\fxd-i\frac{\rtsx}{\p} 
                                     (a_1 - i a_2)^2 (a_0 - i a_3)       \\
  \lan a|{\sst +\thtw, -\thtw, -\thtw} \ran \fxd &=&\fxd- \frac{\rttw}{\p} 
                                     (a_1 - i a_2)^3 
\eea

\section{Wavefunctions on the $SO(3)$ manifold}
\label{a:wfn-so3}

The lowest state in this case has $j=m=n=0$ and is simply a constant
\be \lan R|0,0,0\ran = 1  \;.
\ee
The next higher states are those with $j=1$. There are nine of these. 
Those with $j=1$, $m=1$ 
\bea \lan R|1,+1,+1 \ran \fxd &=& \fxd  
              R_{11}-R_{22} +i(R_{12}+R_{21}) = R^{++}     \nonum \\ \\
     \lan R|1,+1,\;\;\,0 \ran \fxd &=& \fxd 
            -(R_{13}+i R_{23})                =-R^{+0}     \\
     \lan R|1,+1,-1 \ran \fxd &=& \fxd  
            -(R_{11}+R_{22} -i(R_{12}-R_{21}))=-R^{+-} \;, \nonum \\  
\eea
those with $j=1$, $m=0$ 
\bea \lan R|1,\;\;\,0,+1 \ran \fxd &=& \fxd
            -(R_{31} + i R_{32})              =-R^{0+}     \\  
     \lan R|1,\;\;\,0,\;\;\,0 \ran \fxd &=& \fxd
              R_{33}                          = R^{00}     \\
     \lan R|1,\;\;\,0,-1 \ran \fxd &=& \fxd
              R_{31} - i R_{32}               = R^{0-}     \;,  
\eea 
and with $j=1$, $m=-1$ 
\bea \lan R|1,-1,+1 \ran \fxd &=& \fxd
            -(R_{11}+R_{22} +i(R_{12}-R_{21}))=-R^{-+}     \nonum \\ \\ 
     \lan R|1,-1,\;\;\,0 \ran \fxd &=& \fxd
              R_{13} - i R_{23}               = R^{-0}     \\ 
     \lan R|1,-1,-1 \ran \fxd &=& \fxd
              R_{11}-R_{22} -i(R_{12}+R_{21}) = R^{--} \;. \nonum \\ 
\eea 
Higher wavefunctions with total (iso)spin $j$ are made up of products
of $R^{pq}$ to power $j$. The wavefunctions here have not been normalized.

\end{document}